\documentclass[sigconf,nonacm]{acmart} 

\usepackage{soul}

\usepackage{graphicx,comment}
\usepackage{amsmath}
\usepackage{algorithm}
\usepackage{verbatim}
\usepackage{multirow}
\usepackage{booktabs}
\usepackage{makecell}
\usepackage{longtable}
\usepackage{soul}
\usepackage{enumitem}
\usepackage{xspace}
\usepackage{bm}

\usepackage{color, colortbl}
\definecolor{codegreen}{rgb}{0,0.6,0}
\definecolor{codegray}{rgb}{0.5,0.5,0.5}
\definecolor{codepurple}{rgb}{0.58,0,0.82}
\definecolor{backcolour}{rgb}{0.95,0.95,0.92}
\definecolor{LightCyan}{rgb}{0.88,1,1}
\definecolor{LightRed}{RGB}{255, 204, 203}

\usepackage{amsmath}

\newcommand\VRule[1][\arrayrulewidth]{\vrule width #1}


\makeatletter
\def\adl@drawiv#1#2#3{%
        \hskip.5\tabcolsep
        \xleaders#3{#2.5\@tempdimb #1{1}#2.5\@tempdimb}%
                #2\z@ plus1fil minus1fil\relax
        \hskip.5\tabcolsep}
\newcommand{\cdashlinelr}[1]{%
  \noalign{\vskip\aboverulesep
           \global\let\@dashdrawstore\adl@draw
           \global\let\adl@draw\adl@drawiv}
  \cdashline{#1}
  \noalign{\global\let\adl@draw\@dashdrawstore
           \vskip\belowrulesep}}
\makeatother

\usepackage{pifont}

\usepackage{minibox}
\newcounter{observcntr}

\AtBeginDocument{%
  \providecommand\BibTeX{{%
    \normalfont B\kern-0.5em{\scshape i\kern-0.25em b}\kern-0.8em\TeX}}}

\copyrightyear{2023}
\acmYear{2023}
\setcopyright{rightsretained}
\acmConference[WDC '23]{The 2nd Workshop on the security implications of Deepfakes and Cheapfakes}{July 10, 2023}{Melbourne, VIC, Australia}
\acmBooktitle{The 2nd Workshop on the security implications of Deepfakes and Cheapfakes (WDC '23), July 10, 2023, Melbourne, VIC, Australia}
\acmDOI{10.1145/3595353.3595882}
\acmISBN{979-8-4007-0203-7/23/07}

\begin{document}

\title{Why Do Facial Deepfake Detectors Fail?}


\author{Binh Le}
\affiliation{%
  \institution{Sungkyunkwan University}
  \country{South Korea}}
\email{bmle@g.skku.edu}

\author{Shahroz Tariq}
\affiliation{%
  \institution{CSIRO's Data61}
  \country{Australia}}
\email{shahroz.tariq@data61.csiro.au}

\author{Alsharif Abuadbba}
\affiliation{%
  \institution{CSIRO's Data61}
  \country{Australia}}
\email{sharif.abuadbba@data61.csiro.au}

\author{Kristen Moore}
\affiliation{%
  \institution{CSIRO's Data61}
  \country{Australia}}
\email{kristen.moore@data61.csiro.au}

\author{Simon S. Woo}
\affiliation{%
  \institution{Sungkyunkwan University}
  \country{South Korea}}
\email{swoo@g.skku.edu}







\renewcommand{\shortauthors}{Binh Le et al.}



\begin{abstract}
Recent rapid advancements in deepfake technology have allowed the creation of highly realistic fake media, such as video, image, and audio. These materials pose significant challenges to human authentication, such as impersonation, misinformation, or even a threat to national security. To keep pace with these rapid advancements, several deepfake detection algorithms have been proposed, leading to an ongoing arms race between deepfake creators and deepfake detectors. Nevertheless, these detectors are often unreliable and frequently fail to detect deepfakes. This study highlights the challenges they face in detecting deepfakes, including (1) the pre-processing pipeline of artifacts and (2) the fact that generators of new, unseen deepfake samples have not been considered when building the defense models. Our work sheds light on the need for further research and development in this field to create more robust and reliable detectors.
\end{abstract}
\maketitle
\section{Introduction}

\begin{figure}[t]
\centering
\includegraphics[width=0.8\linewidth]{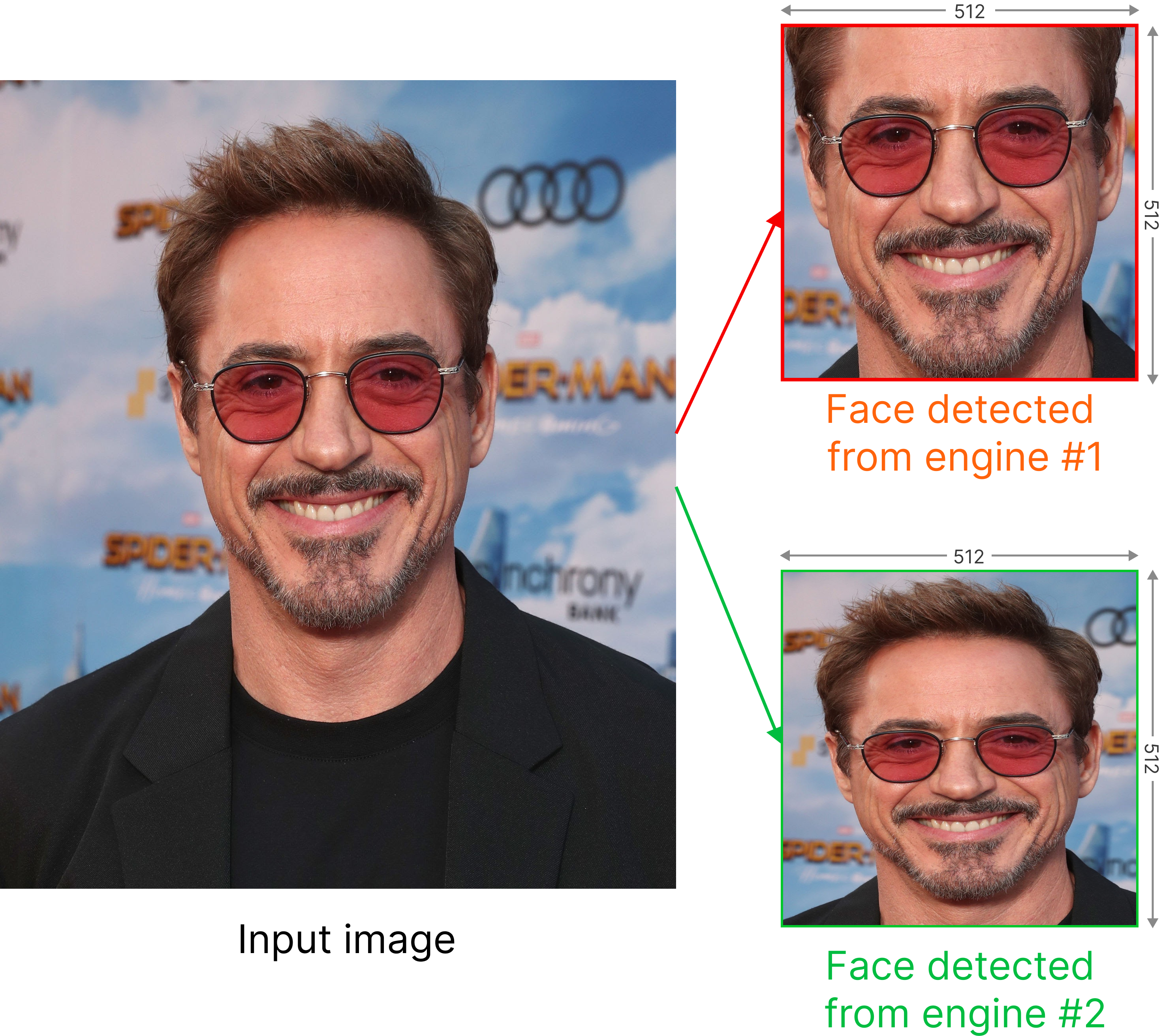}
\vspace{-10pt}
\caption{Illustration of  the face detected from different engines. While the first approach uses a fixed-size central cropped patch from the detected face, the second resizes it, resulting in stretched input image for detectors.}
\vspace{-15pt}
\label{fig:facesize_dist}
\end{figure}



Deepfakes, a term derived from ``deep learning'' and ``fake'' has gained popularity in recent years due to their ability to manipulate images, videos, and audio in a highly realistic manner using artificial intelligence (AI) algorithms. With the increasing sophistication of deepfakes, there is a growing need for effective methods of deepfake detection to combat their potential for harm~\cite{DeepfakeThreat, tariq2022real}. In response, an increasing number of deepfake detection methods have been proposed, employing techniques such as biometric analysis using appearance and behaviors~\cite{AppearanceBehavior}, inter-frame inconsistencies using spatiotemporal data~\cite{CLRNet},  texture enhancement with multi-attention maps~\cite{MultiAttention}, few shot-based~\cite{TAR} and continual learning-based~\cite{Fretal,CoReD} methods for deepfake detection generalization, in addition to other deep learning algorithms~\cite{ShallowNet, le2021exploring, lee2022bznet, bang2021fdftnet}.


\begin{table*}[!t]
\centering
\caption{Performance (ACC and AUC) of deepfake detector trained on raw FaceForensics++ datasets, and validated under different circumstances.} 
\label{tab:exp_results}
\vspace{-10pt}
\begin{tabular}{l !{\VRule} c  !{\VRule} c  !{\VRule} c c  !{\VRule}  c !{\VRule} c  }
    \hline
    Test type & FF++ test set (dlib$_4$)& Video comp. & dlib$_{15}$ & MTCNN$_4$ & Adv. noise & CelebDF-v2 \\
    \hline
    ACC & \textbf{0.980} & 0.615 (\textcolor{red}{$\downarrow$ .365}) & 0.951 (\textcolor{red}{$\downarrow$ .290}) &0.970 (\textcolor{red}{$\downarrow$ .010}) &0.002 (\textcolor{red}{$\downarrow$ .978}) & 0.526 (\textcolor{red}{$\downarrow$ .454})\\
    AUC &  \textbf{0.994} &0.788 (\textcolor{red}{$\downarrow$ .206}) & 0.989 (\textcolor{red}{$\downarrow$ .005}) & 0.993 (\textcolor{red}{$\downarrow$ .001}) & 0.000 (\textcolor{red}{$\downarrow$ .994}) & 0.573 (\textcolor{red}{$\downarrow$ .421}) \\
    \hline
\end{tabular}%
\end{table*}

The robustness of deepfake detectors is crucial, particularly in cybersecurity applications such as Facial Liveness Verification, where their failure could have serious consequences~\cite{li2022seeing}. In this paper, we aim to shed light on some of the challenges that deepfake detectors face when deployed in real-world situations. By exploring these obstacles, we hope to provide insight into the limitations of current deepfake detection methods and stimulate further research in this critical area.

{To address the challenges of deepfake detection, it is crucial to understand the limitations and pitfalls of existing approaches. Despite the emphasis on detection accuracy, there is a lack of consideration for \textit{explainability} in deepfake detection. In this article, we will delve into two specific scenarios where using deepfake detectors may lead to unexpected results. Firstly, a mismatch between the pre-processing pipeline used in the deployment and the one used during training can compromise the detector's performance. Secondly, a lack of diversity in the datasets used during training can lead to biased and unreliable results.}

\textbf{Pre-processing: }
A newcomer to the field of deepfake detection may encounter challenges in obtaining accurate results when using an off-the-shelf detection tool, as they may not be familiar with the pre-processing pipeline of the tool. It is important to note that deepfake videos are not simply fed directly into the detector. This is because the deepfake detection models typically require the input of a certain size, whereas the original image/video may be a different size. To deal with this, pre-processing is performed, which may obscure the deepfake artifacts that the detector relies on. Worse, there is no standard pre-processing pipeline nor a standardized size for inputs to deepfake detectors. This paper aims to shed light on the impact of pre-processing on the detection process and provide explainability/guidance around one of the major preprocessing tasks --- when to crop versus resize input.


Pre-processing techniques, such as resizing and cropping, are widely used in the pipeline of deepfake detectors, but their effects on detection performance are often overlooked. Our analysis reveals that resizing elongates the face to match the specified size, while cropping does not have this effect. This can lead to issues if the model is trained on stretched faces but deployed on naturally proportioned images. On the other hand, shrinking the input size through resizing may result in crucial features being lost. Therefore, we have found that selecting resizing over cropping when reducing the input size would negatively impact detection accuracy.


\textbf{Dataset/Deepfake generator diversity: }
In recent years, the diversity of deepfake datasets and deepfake generators has proliferated, supporting research in the area. Some of the most popular deepfake datasets include DeepFake Detection Challenge \cite{dolhansky2020deepfake}, FaceForensics++ \cite{rossler2019faceforensics++}, and Celeb-DF \cite{li2020celeb}; and they vary in terms of quality and methods used for generating them. Alongside this, a wide range of generators have been published, aiming to be more accessible and user-friendly, including DeepFaceLab, GAN-based and AutoEncoder-based generators.

Nevertheless, most state-of-the-art (SoTA) deepfake detectors have been developed to detect a specific type of deepfake dataset, leading to performance degradation on new, unseen deepfakes. Their generalization limitations also include the variation in deepfake quality, as demonstrated by their poor performance on compressed or manipulated inputs. Our study illuminates the correlation between deepfakes datasets. In particular, we employ both frequency transformation and deep-learning embeddings to visualize their interdependence and distribution. In this way, we highlight undisclosed reasons that may lead to the poor performance of a biased detector that was learned from limited types of deepfakes or limited generation toolkits.


\section{Background}
While the term deepfakes can be used to refer to any artificial replacement using AI, we limit ourselves in this work to facial deepfakes~\cite{mirsky2021creation}. In general, there are two categories of facial manipulation approaches: face reenactment and face-swapping. Face reenactment involves changing the facial expressions, movements, and speech of a person in a video to make it appear as though they are saying or doing things they never actually did. In face-swap deepfakes, the face of a person in a video or image is replaced with someone else's face, making it appear as if the latter person was present in the original footage.

In this study, we utilize the FaceForensics++ dataset \cite{rossler2019faceforensics++}, which is a well-known deepfake dataset that was created for validating different deepfake detection algorithms. From 1000 real videos, the authors generated a corresponding 1000 synthesized videos using DeepFakes \cite{deepfakes}, Face2Face \cite{thies2016face2face}, FaceSwap \cite{faceswap}, NeuralTextures \cite{thies2019deferred}, and FaceShifter \cite{li2019faceshifter} algorithms. Among these, NeuralTextures and Face2Face are reenactment methods; the others are face-swapping algorithms. In addition, to increase the diversity, we include CelebDF-v2 \cite{li2020celeb} dataset, which is created by several published deepfake apps for face swapping, and fine-tuned by a sequence of post-processing steps, making it a highly realistic dataset.

\section{Methodology}
\subsection{Data-preprocessing}
 For FaceForensics++ datasets, we follow the same preprocessing step as in ADD \cite{le2021add}: 92,160, 17,920, and 17,920 images for training, validation, and testing, respectively. Each set has a balanced number of real and fake images, and the fake images are derived from all five deepfake datasets. For the CelebDF-v2, we used 16,400 for solely validating the pre-trained model.
 
 In order to detect faces from a video, we used the dlib~\cite{dlib09} toolkit with padding factor of $15\%$ and $3\%$, respectively, and MTCNN with padding of 4\% to simulate different face detection engines.

 \subsection{Training and validation}
 We utilized ResNet50 as our detector and built a binary classifier. All the input images were resized to $224 \times 224$, and we used detected faces with padding of $3\%$ for training. The models were trained with the Adam \cite{kingma2014adam} optimizer with a learning rate of $2e-3$, scheduled by one-cycle strategy \cite{smith2019super}. Only random horizontal flip is applied during training. We used a mini-batch size of 192. During every epoch, the model was evaluated ten times, and we saved the best weight based on the validation accuracy. Early stopping \cite{prechelt1998early} was applied when the model didn't not improve after $10$ consecutive validation times.

\section{Evaluation}

\begin{figure*}[t]
\includegraphics[width=0.495\linewidth]{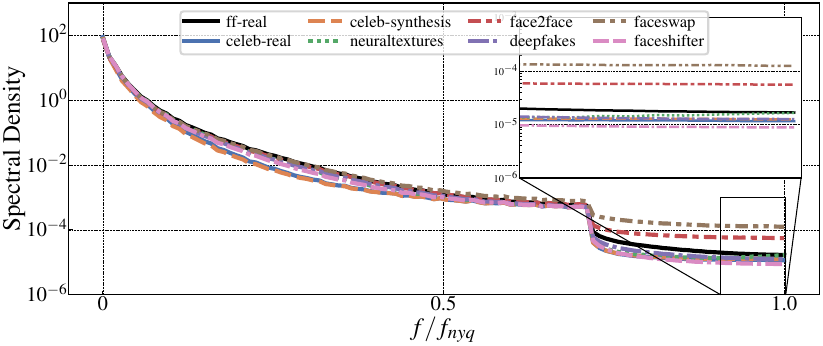}\hfill
\includegraphics[width=0.495\linewidth]{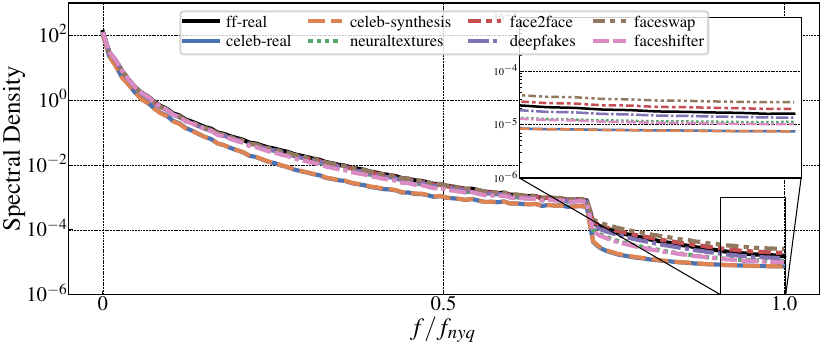}
\vspace{-10pt}
\caption{High-frequency discrepancies of central cropped face (Left; Crop: 156) vs. resize large cropped face (Right; Resize: 156).}
\label{fig:crop_vs_resize}
\end{figure*}

\begin{figure*}[t]
\centering
\includegraphics[clip, trim=10pt 10pt 55pt 10pt,width=0.495\linewidth]{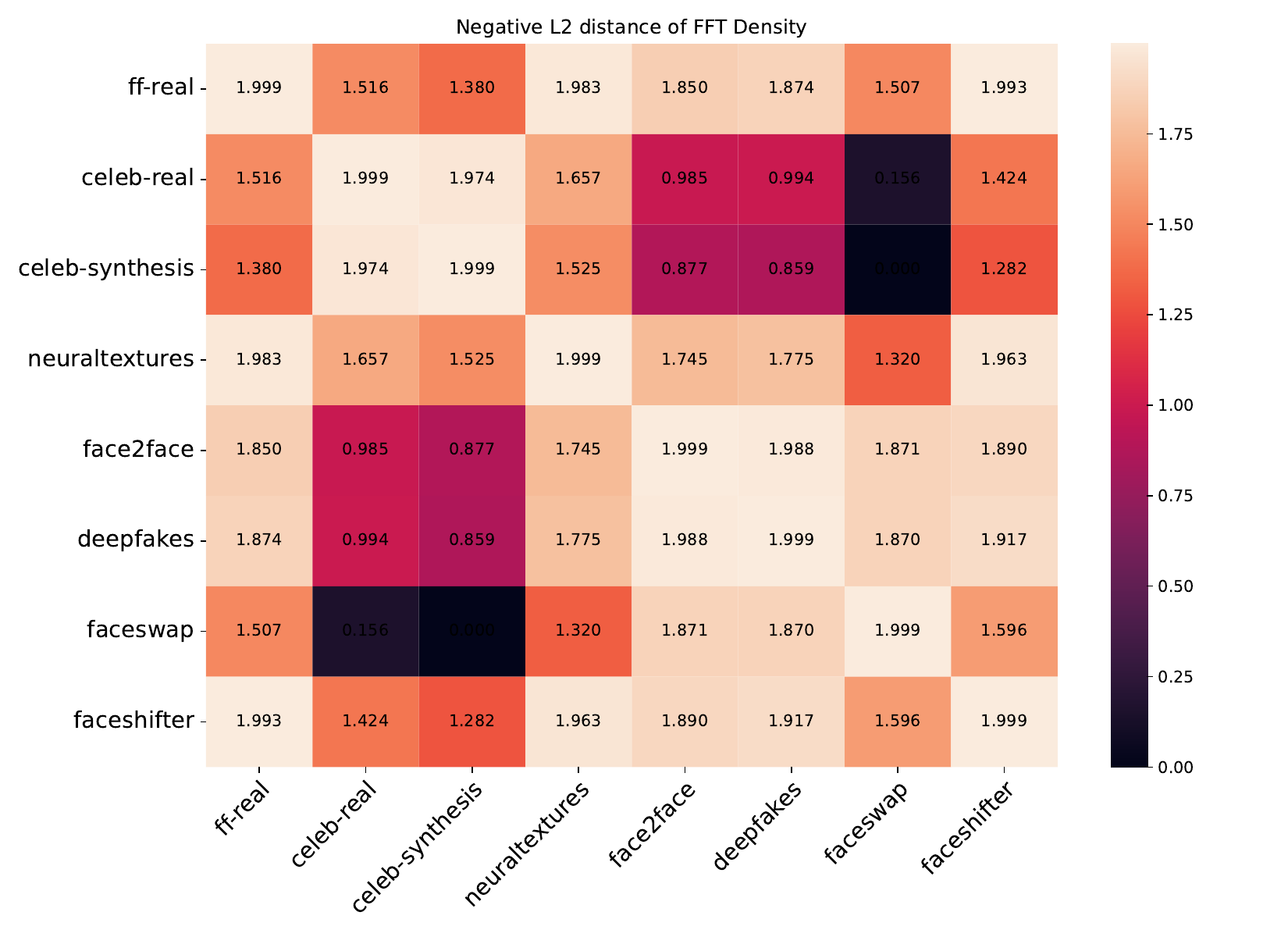}\hfill
\includegraphics[clip, trim=10pt 10pt 55pt 10pt,width=0.495\linewidth]{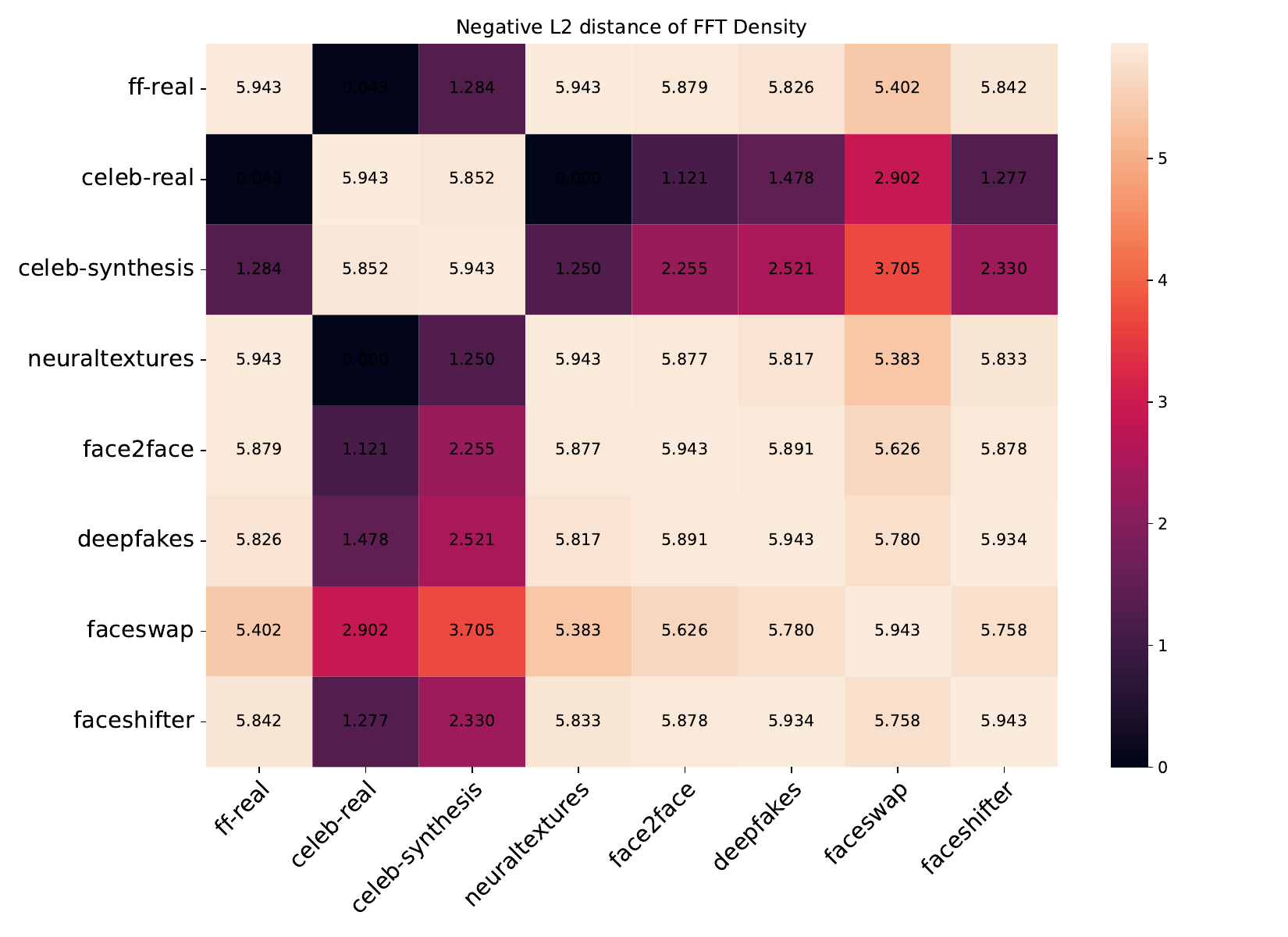}
\vspace{-10pt}
\caption{Similarities between frequency density lines from Figure \ref{fig:crop_vs_resize}. The higher values indicate the higher similarities between datasets (Left; Crop: 156 and Right; Resize: 156).}
\label{fig:heatmap_l2}
\end{figure*}

\begin{figure}[t]
\centering
\includegraphics[clip, trim=35pt 45pt 57pt 69pt,width=1\linewidth]{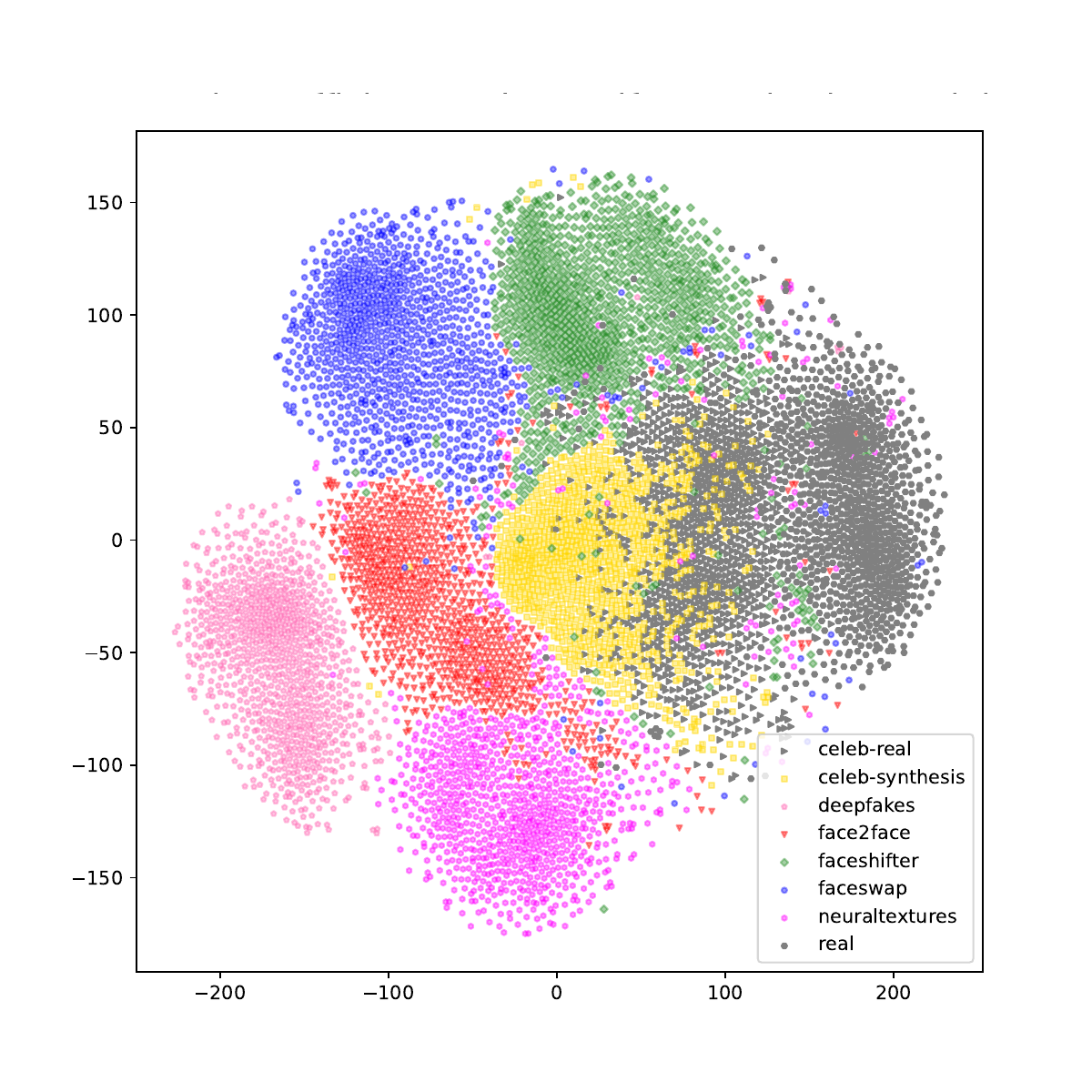}
\vspace{-10pt}
\caption{t-SNE visualization of various deepfake datasets using a pre-trained self-supervised learning embedding model}
\label{fig:data_dist}
\end{figure}

In this section, we evaluate the pre-trained model under the following different scenarios discussed below. The experimental results are presented in Table \ref{tab:exp_results}.
\subsection{Video compression} Deepfakes are detected by their subtle artifacts, represented by high-frequency components. Various methods of lossy compression, including video compression and JPEG compression, can successfully eliminate these fine-grained artifacts, leading high rate of false-positive prediction. As shown in the second column of Table \ref{tab:exp_results},  we applied the H.264 codec with constant rate quantization parameters of 23 to raw videos. As a result, the pre-trained detector drastically dropped its accuracy from 98\% to 61.2\%.

\subsection{Face extraction approach} 
The effects of using different face extraction approaches are indicated in the fourth and fifth columns of Table  \ref{tab:exp_results}. While the MTCNN detector can slightly reduce the performance of the ResNet50 model, using dlib with larger padding can substantially decrease its performance in terms of both accuracy and AUC scores. We argue that since the model had learned only from the facial features, the complex background in some contexts affected the model's attention. 

Our second exercise in this category was inspired by a recent live-face detection algorithm \cite{wang2022patchnet} which proposed to use fixed-size patches cropped from the original faces instead of resizing the input. The explained reason is that resizing step can distort the discriminative features. To further examine this hypothesis, we conducted a pilot study in which we selected 5,000 images from each deepfake dataset and performed central crop and resizing steps, respectively, on them, as illustrated in Figure \ref{fig:facesize_dist}. Next, we applied Fourier transformation and extracted the average density representation of each dataset in the frequency domain \cite{durall2020watch}. The results are provided in Figure \ref{fig:crop_vs_resize}. As one may observe, central cropping results in high-frequency differences between datasets. Resizing, on the other hand, pushes the high-frequency representations of datasets close together, making it difficult to distinguish.

\subsection{Adversarial noise}
Deep neural networks are well known for their adversarial nature showing through their vulnerability against adversarial examples. The adversarial samples are created by adding small, often imperceptible, perturbations to the original inputs. To validate this property, we apply one-step $L_\infty$ white-box PGD attack \cite{madry2017towards} with a small perturbation size of $1/255$ and step size $\epsilon=1/255$. As indicated in the sixth column of Table \ref{tab:exp_results}, almost all the predictions are flipped, as demonstrated by an accuracy score close to \textit{zero}. Therefore, deploying deepfake detectors in practice should consider this aspect and have proper pre-processing steps or defense mechanisms to eliminate the effect of adversarial samples.

\subsection{Data shift}
Data shift refers to changes in the statistical properties of the data distribution used to train the detection model compared to the distribution of data the model encounters in deployment. In fact, data shift in deepfake detection can be a result of different factors: ethnicity (\textit{e.g.}, Asian vs. African), environment (\textit{e.g.}, indoor vs. outdoor), generating method (\textit{e.g.}, Neural texture vs. FaceSwap). We show the results of cross-dataset validation in the final columns of Table \ref{tab:exp_results}. Although the model was trained over five datasets of FaceForensics++, it still struggles to distinguish deepfake from the CelebDF-v2 dataset, indicated by its performance of approximately random guesses.

To explain this phenomenon, we perform two experiments to visualize the relationships between datasets. First, from the density representations of datasets from Figure \ref{fig:crop_vs_resize}, we use negative distance to indicate the closeness between datasets that are formulated as $max-||a-b||^2_2$. As we can observe from Figure \ref{fig:heatmap_l2}, the cropping step introduces less relationship between datasets compared to resizing. Nevertheless, in both approaches, there is less relation between FaceForensics++ datasets and CelebDF-v2, both in real and deepfake parts. In our second experiment, we utilize a pre-trained ``self-supervised learning'' model, SBI \cite{shiohara2022detecting}, with EfficientNet-B4 backbone to get the intermediate representations of each deepfake dataset. As illustrated in Figure \ref{fig:data_dist}, each deepfake dataset has its own distribution in the latent space. Therefore, if a detection model solely learns a single dataset, its decision boundary may lose its generalization for others, leading to the degradation of its performance.

\section{Remarks}

Despite a plethora of ongoing research aimed at improving the accuracy of deepfake detectors, there is also a multitude of factors that hinder their performance. These include pre-processing steps, intended manipulation from attackers, and ongoing advancement of deepfake technology induces the low generalization of pre-trained detectors. In this paper, we quantially and visually expose these factors from the explainability viewpoints. This study also raises the awareness of researchers of not only developing effective deepfake detectors but also putting their effort into mitigating those crucial factors, reducing false positive and negative rates in practice.

\begin{acks}
This work was partly supported by Institute for Information \& communication Technology Planning \& Evaluation (IITP) grants funded by the Korean government MSIT: (No. 2022-0-01199, Graduate School of Convergence Security at Sungkyunkwan University), (No. 2022-0-01045, Self-directed Multi-Modal Intelligence for solving unknown, open domain problems), (No. 2022-0-00688, AI Platform to Fully Adapt and Reflect Privacy-Policy Changes), (No. 2021-0-02068, Artificial Intelligence Innovation Hub), (No. 2019-0-00421, AI Graduate School Support Program at Sungkyunkwan University), and (No. 2023-00230337, Advanced and Proactive AI Platform Research and Development Against Malicious Deepfakes).
\end{acks}

 
\bibliographystyle{ACM-Reference-Format}
\bibliography{ref}
\end{document}